\documentclass[10pt, conference, a4paper]{IEEEtran}

\usepackage{booktabs} 
\usepackage{balance}
\usepackage[ruled]{algorithm2e}
\usepackage[]{xcolor}
\usepackage{tabularx}
\usepackage{graphicx}
\usepackage{amsmath}
\usepackage{amssymb}
\usepackage{tikz}
\usetikzlibrary{matrix}
\usepackage{multirow}
\usepackage{float}
\usepackage{soul}
\usepackage{svg}
\newtheorem{definition}{Definition}

\ifCLASSOPTIONcompsoc
  \usepackage[nocompress]{cite}
\else
  \usepackage{cite}
\fi

\begin{document}

\title{The Limits of LoRaWAN in Event-Triggered\\Wireless Networked Control Systems}
\author{\IEEEauthorblockN{Ivana Tomi\'c,
Laksh Bhatia, Michael J. Breza and
Julie A. McCann}
\IEEEauthorblockA{Department of Computing, Imperial College London, UK\\
Email: \{i.tomic, laksh.bhatia16, michael.breza04, j.mccann\}@imperial.ac.uk}}
\maketitle

\markboth{Journal of \LaTeX\ Class Files,~Vol.~14, No.~8, August~2015}%
{Shell \MakeLowercase{\textit{et al.}}: Bare Advanced Demo of IEEEtran.cls for IEEE Computer Society Journals}

\IEEEtitleabstractindextext{%
\begin{abstract}

Wireless sensors and actuators offer benefits to large industrial control systems.
The absence of wires for communication reduces the deployment cost, maintenance
effort, and provides greater flexibility for sensor and actuator location and system
architecture.
These benefits come at a cost of a high probability of communication delay or
message loss due to the unreliability of radio-based communication.
This unreliability poses a challenge to contemporary control systems that are designed
with the assumption of instantaneous and reliable communication. 
Wireless sensors and actuators create a paradigm shift in engineering energy-efficient
control schemes coupled with robust communication schemes that can maintain system
stability in the face of unreliable communication.
This paper investigates the feasibility of using the low-power wide-area communication
protocol LoRaWAN with an event-triggered control scheme through modelling in Matlab.
We show that LoRaWAN is capable of meeting the maximum delay and message loss
requirements of an event-triggered controller for certain classes of applications.
We also expose the limitation in the use of LoRaWAN when message size or communication
range requirements increase or the underlying physical system is exposed to significant
external disturbances.

\end{abstract}

}

\maketitle

\IEEEdisplaynontitleabstractindextext

\IEEEpeerreviewmaketitle

\ifCLASSOPTIONcompsoc
\IEEEraisesectionheading{\section{Introduction}\label{sec:introduction}}
\else
\section{Introduction}
\label{sec:intro}
\fi

The design of traditional periodic control systems \cite{Astrom1997} assumes
instantaneous and reliable communication of sensor data and control actions 
between the sensors and the controllers.
This assumption is easy to meet when the sensors and controllers are connected
by cables or wires.
The controller will have relevant data to maintain the stability and the 
desired level of performance of the controlled physical system.
Periodic transmissions at a high frequency can be achieved because wired 
systems do not pose restrictions on the bandwidth available.

There is a current movement to instrument industrial control systems like
water/waste distribution, unmanned off-shore oil rigs, transportation networks,
and agricultural facilities with wireless sensors and actuators.
Wireless Networked Control Systems (WNCSs) are smarter, more responsive to user
demand, and more efficient in their use of resources such as energy 
\cite{Rajkumar2010}.
These next generation industrial systems are composed of a network of sensor 
nodes installed around a plant to measure its physical processes and transmit
the measurements (data) via a wireless network.
The controller receives the measurements and processes them according to 
an underlying control scheme.
The result is a control action that is sent wirelessly to actuators to influence
the dynamics of the plant.
The challenge of this approach is that wireless radio networks are not reliable,
and the assumption of instantaneous and reliable communication no longer holds 
\cite{Ali2015}.

There are two associated challenges with WNCSs. The first is to develop control
schemes that can tolerate a certain amount of delay and data loss caused by 
unreliable wireless networks.
The second is to develop wireless network protocols that can provide the bounded
maximum delay and message loss required by the control scheme.
To provide the solution that addresses both challenges, the communication and
control systems have to be jointly designed.
This is known as the communication and control systems co-design problem 
\cite{Park2017, Wu2017}.
An additional constraint on the communication and control systems design problem
is energy efficiency.
WNCS sensors are battery powered which introduces restrictions on the available
bandwidth.
They are often deployed in inaccessible locations which makes frequent battery
changes difficult.
In this paper, we investigate the feasibility of using the low-power wide-area 
communication protocol LoRaWAN with an event-triggered control scheme.

Event-Triggered Control (ETC) schemes \cite{Tabuada2007} are a solution to the
high communication cost and frequent battery changes of traditional periodic
control schemes.
Traditional periodic control schemes send communication every fixed period of
time, even if there is no change in the underlying physical process.
ETC schemes save energy by only sending communication when an event occurs
and new action is needed.
An event is triggered when there is an indication that the stability or 
performance of the system are about to be compromised.

Low-Power Wide-Area (LPWA) networks \cite{Raza2017} have been developed to 
enable wireless communication over long ranges.
LPWA techniques enable long-range communication of up to $15$km's at a low 
data rate of $0.3$-$37.5$Kbps.
This offers trade-off between communication coverage, and data rates when 
compared to commonly used short-range protocols such as those used by 
WirelessHART and ISA-100.a.

In this paper we provide the following contributions:
\begin{itemize}
\item We model the delays and message losses introduced by
the LPWA communication protocol LoRaWAN \cite{LoRaWAN2015} for different
rates and message sizes.
We analyse an application scenario where we model a linear ETC system in 
Matlab. 
We evaluate the effects of LoRaWAN delays and message loss rates on system
stability and performance guarantees for both cases, an ideal ETC system
and the ETC system that is exposed to external disturbances.
\item Most of the existing studies that address event-triggered controllers
with radio networks assume that bounds on communication delays and packet
losses are given a priori (e.g. \cite{Wang2011, Lehmann2012}).
However, it is difficult to obtain such properties in real applications.
In this paper we consider a practical communication protocol LoRaWAN.
LoRaWAN is an example of a long-range protocol whose use in control 
scenarios has received very little attention \cite{Park2017}. 
\end{itemize}

The rest of the paper is organized as follows: Sec.~2 presents the 
problem formulation.
Sec.~3 presents the event-triggered control model.
Sec.~4 presents the LoRaWAN communication model.
We give the evaluation results in Sec.~5 and end the paper in Sec.~6
with brief concluding remarks.

\section{Wireless Networked Control System Problem Formulation}

\begin{figure}[!t]
\begin{center}
\includegraphics*[width=0.9\linewidth]{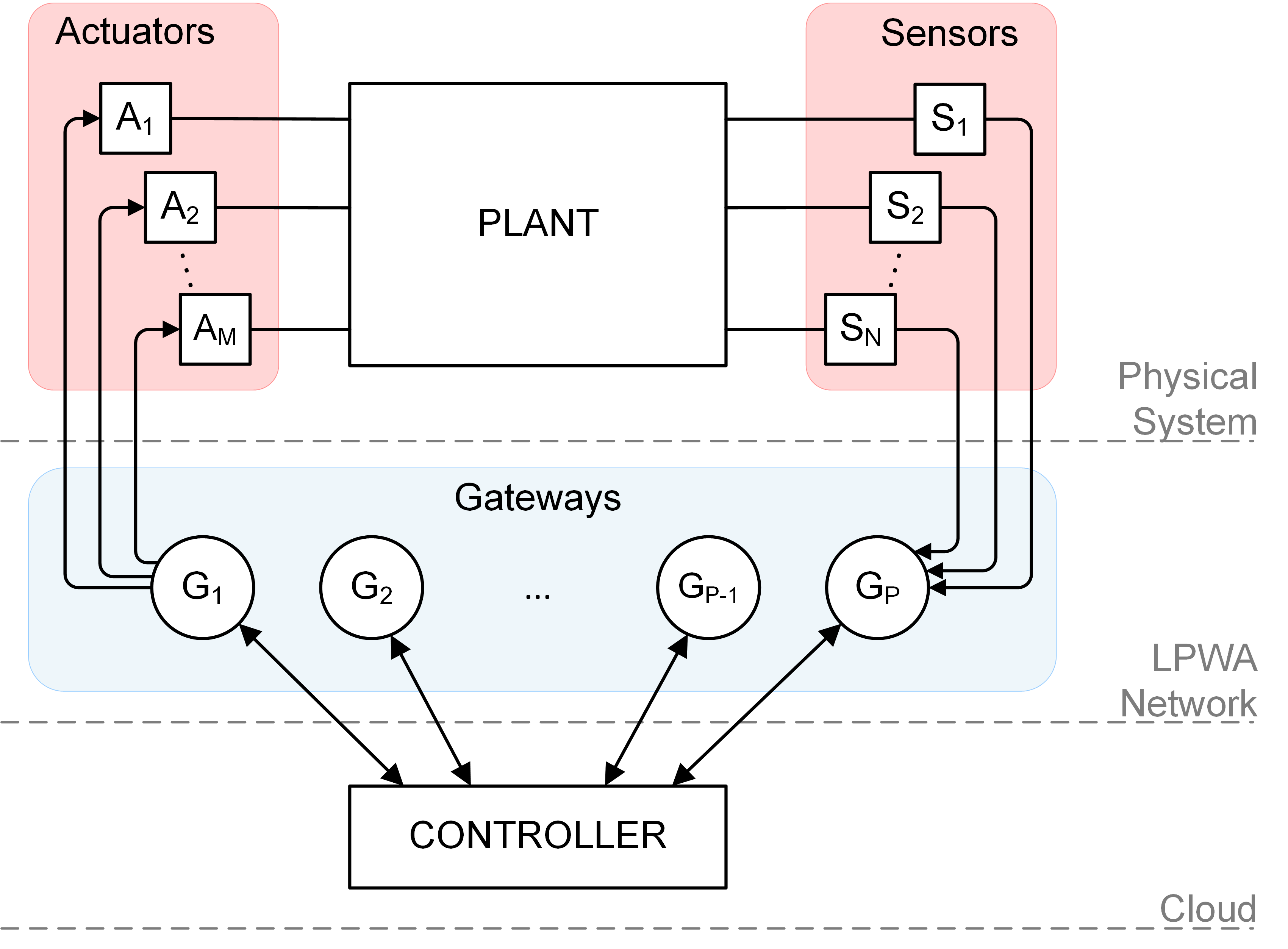}
\caption{\label{fig:architecture} The architecture of wireless networked control system.}
\vspace{-5mm}
\end{center}
\end{figure}

In Fig.~\ref{fig:architecture} we present a diagram of the closed-loop
WNCS considered in this paper.
The WNCS consists of a large complex physical process (the plant) and 
the control and management system which are connected via a LPWA 
communication network.
The plant is a continuous-time physical system instrumented with the
set of sensors $\{\mathcal{S}_1, \mathcal{S}_2, \ldots, \mathcal{S}_N\}$
and the set of actuators $\{\mathcal{A}_1, \mathcal{A}_2, \ldots, \mathcal{A}_M\}$.
We assume that all of the sensors and actuators are collocated on 
LPWA-enabled end-devices which share same communication channels.

The end-devices communicate with the set of gateways 
$\{\mathcal{G}_1, \mathcal{G}_2, \ldots, \mathcal{G}_P\}$ when a change in
the plant is detected.
Both, sensor-to-gateway and gateway-to-actuator communication is achieved 
in a single-hop fashion via a LPWA network.
A single-hop network topology is far more favourable for control-based 
systems due to its high reliability and low energy cost compared to multi-hop
networks.
The information exchange between gateways and the controller is achieved via
traditional wired communication and thus is instantaneous and reliable.
In the rest of this paper the terms the gateway and the controller will be 
used interchangeably.

To describe the WNCS in Fig.~\ref{fig:architecture} we present two models:
\begin{enumerate}
\item \emph{Event-Triggered Control Model} -
A general model for event-triggered control of a linear plant.
\item \emph{LoRaWAN Communication Model} -
We explain how LoRaWAN works and present its main concepts and parameters. 
\end{enumerate}
%

We use the notion of communication delays and message losses to investigate
the feasibility of using LoRaWAN with an event-triggered control system.
The LoRaWAN introduces delays and losses as part of the protocol.
In an ETC system there is a maximum bound on the delay and data loss for
which the system can remain stable and a certain level of performance is
guaranteed.
In this paper, we check if these bounds can be met by LoRaWAN.

\section{Event-Triggered Control Model}
\label{sec:sys_model}
Here we give a model of our event-triggered control scheme as a linear control
system of the form:
\begin{align} \label{eq:lin_sys}
\dot{\xi}(t)&=A\xi(t)+B\upsilon(t), \quad \xi(0)=\xi_0
\end{align}
where $\xi(t) \in \textrm{R}^n$ denotes the state input from the sensors
and $\upsilon(t) \in \textrm{R}^m$ denotes the control action sent to the 
actuator at time $t$.
The matrix $A \in \textrm{R}^{n\times n}$ is the state matrix and 
$B \in \textrm{R}^{n\times m}$ is the input matrix.

The system is connected via a communication channel to the controller.
In this paper, the communication channel uses LoRaWAN wireless protocol.
In a conventional discrete-time state feedback control, 
the control input $\upsilon(t)$ is given by
\begin{align} \label{eq:feedback_gain}
\upsilon(t)&=K\xi(t_k), \quad \forall t \in [t_k,t_{k+1}]
\end{align}
where $t_k$, $k \in \mathrm{N}$ are the sampling instants which occur
periodically, i.e. $t_k=kh$ for $h>0$.
The feedback gain matrix is denoted as $K\in \textrm{R}^{n\times n}$.

In ETC schemes, the sampled state taken from the sensor, $\xi(t_k)$,
is used to evaluate a predefined triggering condition at each $t_k=kh$
for $k \in \mathrm{N}$, $h>0$.
In that way, the controller only updates and sends a control action
when the triggering condition is satisfied.
We refer to this as to an event and it can be expressed in terms of the
measurement error $\epsilon(t_k)=|\xi(t_k)-\xi(t)|\leq \eta$ for $t \in
[t_k,t_{k+1}]$ that exceeds the predefined threshold value $\eta$.
This approach differs from conventional periodic control schemes where
events are transmitted regularly regardless of the state of the plant.
ETC schemes only transmit an event if one actually occurs.
This difference reduces the computational load of the controller and
the amount of communication required between the sensors/actuators and
the controller.
The reduction of computation and communication equate to direct energy
savings for the control system.
This energy saving aspect is why ETC strategies are actively researched
for WNCSs.

In general, the controller is centralized and it works in 
a sample-and-hold fashion:
\begin{align} \label{eq:c1_update}
\upsilon(t_k)=\left\{\begin{array}{l l}
    K\xi(t_k), \quad \quad \textrm{if} \quad \theta_k=1 \\
    K\xi(t_{k-1}), \quad \textrm{if} \quad \theta_k=0
   \end{array}\right. .
\end{align}
The indicator function $\theta_k=1$ indicates that the triggering
condition is satisfied, $\theta_k=0$ indicates that the triggering
condition is not satisfied at time $t_k$.
Satisfaction of the triggering condition is based on measurements
received from each end-device.

With ETC control schemes in \cite{Mazo2014} the event-triggering
mechanism is distributed to the end-devices such that
$j\in\{1,\ldots, N_d\}$, where $N_d$ is the total number of end-devices
in the network.
End-devices work asynchronously.
Only the data of the end-device that measured a threshold violation is
sent to the gateway to update the control action in Eq.~\ref{eq:c1_update},
i.e. if $j\in\{1,\ldots,N_d\}$ and $\theta_{k_j}=1$, then
$\xi(t_k)=[\xi_1(t_{k-1}), \xi_2(t_{k-1}), \ldots, \xi_j(t_k), \ldots,
\xi_N(t_{k-1})]^T$.
The decentralisation of the event triggering mechanism further increases
energy efficiency by reducing the number of events.

There exists a non-zero minimum time that must always elapse after an
event and before the next event is triggered.
This lower bound on inter-arrival time can be explicitly computed 
(see \cite{Tabuada2007}).
We show later that when an ETC system relies on a LoRaWAN,
there is a practical bound imposed by LoRaWAN on the event rate due
to the LoRaWAN protocol and the limitations of wireless communication.
This practical bounds imposed by LoRaWAN can be abstracted to event
delays, control action delays and event losses in the previously
discussed ETC implementation.
We define these as:
\begin{definition}
We denote by $\tau_k^{j\rightarrow C}$ the event delay of the
measurement $\xi(t_k^j)$ of the end-device $j$ to the controller
$C$ at time $t_k^j$.
Similarly, $\tau_k^{C\rightarrow j}$ is the control action delay of
the control action to the end-device $j$.
\end{definition}
\begin{definition}
We denote by $P_k^{i\rightarrow C}$ the number of successive event
losses in the transmission of the measurement $\xi(t_k^j)$ of end-device
$j$ to the controller $C$ at time $t_k^j$.
A loss of an event message also means the loss of the subsequent
control action message as it is dependant on the successful reception
of the event message by the controller.
\end{definition}

Event or control action delays or losses directly affect the ability
of the controller to maintain a stable plant.
They cause the absence of control actions.
The stability or level of performance of the plant is determined by
$\eta$, the size of the error that can be tolerated.
There is a maximum number of consecutive event losses and a maximum
tolerable delay for which the performance of the system can be guaranteed
(see \cite{Mazo2014,Guinaldo2012}).
In the next section we discuss LoRaWAN protocol, and its delays and losses.

\section{LoRaWAN Communication Model}
\label{sec:lora}

In this section we explain how LoRaWAN functions and discuss some
of its parameters.
We relate its parameters and performance to the event and control action delays
and losses discussed in the previous section.


\subsection{Network Architecture and Transmission Parameters}

LoRaWAN provides bidirectional communication with an uplink (end-device to
gateway) and a downlink (gateway to end-device).
LoRaWAN defines three classes of end-devices, classes A, B and C. 
Class A end-devices only receive downlink communication from the gateway
after a successful uplink sent from the end-device to the same gateway.
There are two time periods (or windows) that are available to the gateway
for downlink transmissions at $1$s and $2$s.
Class B end-devices allow the gateway to schedule downlink communication
windows without a prior successful uplink transmission from an end-device.
Class C end-devices listen for transmissions from the gateway all the time
unless they are transmitting.
In this paper, we consider only Class A end-devices.
Class A devices are the most energy efficient, and are a perfect fit for
ETC because communication from the gateway is always triggered by an uplink
transmission from the end-device.

LoRaWAN end-devices transmit uplink packets using an ALOHA-based channel access
scheme. 
The LoRa physical layer uses Chirp Spread Spectrum (CSS) modulation.
CSS signals are modulated by pulses that increase or decrease in frequency,
or chirps.  
The number of chirps used to encode each symbol is given by $2^{\mathrm{SF}}$
where $\mathrm{SF}$ represents the Spreading Factor that varies between $7$
and $12$ in increments of $1$. 
SF7 provides a data rate of $5.468$kbps, while SF12 provides $0.293$kbps.
The data rate in LoRa also depends on the channel bandwidth (we use $125$kHz)
and the code rate (we use $4/5$).
The trade-off in LoRa is that the higher SF has a lower data rate, but a longer
range and is more resilient to interference.
Lower SF has a higher data rate, but a shorter range and less resilience. 

In LoRa, the use of retransmissions is optional.
The use of CSS makes the signals very robust to interference, and it has been
experimentally shown that LoRA is very reliable~\cite{cattani2017experimental}.
If retransmissions are used, the number of and timing is at the discretion of the
each end-device.
In this paper, we assume no use of retransmissions. 

\subsection{Round Trip Time Delay of LoRaWAN}

\begin{figure}[!t]
\begin{center}
\includegraphics*[width=0.83\linewidth]{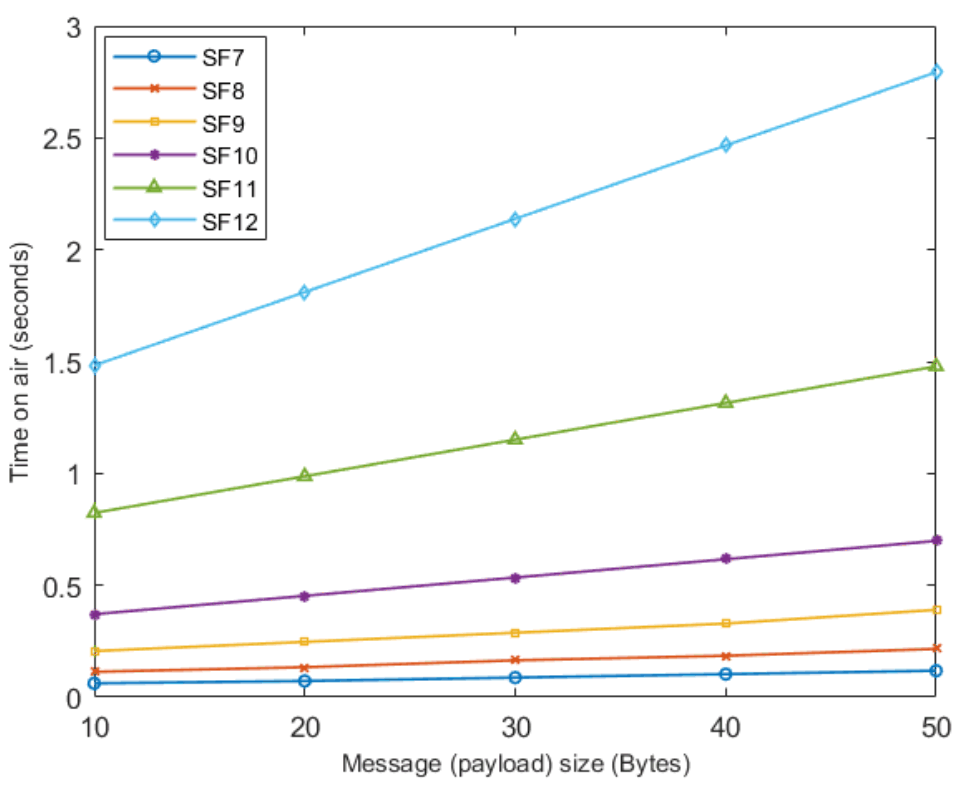}
\caption{\label{fig:timeonair}$\mathrm{TimeOnAir}$ of LoRaWAN for a code rate of $4/5$ and a bandwidth of $125$kHz.}
\vspace{-5mm}
\end{center}
\end{figure}

Wireless technologies such as IEEE 802.15.4 have transmission times in
the range of $10$-$100$ms depending on the payload size \cite{Gutierrez2001}.
LoRaWAN networks have transmission times from $61$ms-$2.7$s depending on the
message size, as it can be seen in Fig.~\ref{fig:timeonair}.
We define the time needed for a message (sensor measurements) to be transmitted
from an end-device to the controller (an uplink transmission) as
$\mathrm{TimeOnAir}$.
Figure~\ref{fig:timeonair} shows the $\mathrm{TimeOnAir}$ of an uplink
transmission using LoRaWAN for a code rate of $4/5$, bandwidth of $125$kHz,
all SF$7$ through $12$ inclusive, and messages sizes of $10$, $20$, $30$, $40$,
and $50$Bytes of payload.
The messages sizes are shown without the header.
The message header adds $13$Bytes regardless of SF.
As can be seen from Fig.~\ref{fig:timeonair}, large SFs increase the
$\mathrm{TimeOnAir}$.
Large SFs also have an impact on duty cycling (channel availability) as shown
later.

Downlink transmissions also use the LoRaWAN protocol.
These transmissions contain the control action sent from the controller to the
actuators connected to the end-devices as in Fig.~\ref{fig:architecture}.
With class A end-devices, downlink transmissions are possible only after
a successful uplink transmission.
We assume that the downlink payload size is the same as the uplink payload size
and both have the same $\mathrm{TimeOnAir}$.

In Def.~1 we defined event delays as $\tau_k^{j\rightarrow C}$ and control
action delays as $\tau_k^{C\rightarrow j}$.
These delays are a result of the used SF.
They are expressed in our LoRaWAN model as $\mathrm{TimeOnAir}$. 
Additionally, the LoRaWAN specification \cite{LoRaWAN2015} introduces a
fixed time delay between an uplink and a downlink transmission.
This delay is $\mathrm{RECEIVE\_DELAY}$ and it equals $1$s.

We now define Round Trip Time ($\mathrm{RTT}$) as:
\begin{equation} \label{eq:RTT}
\mathrm{RTT} = 2\times\mathrm{TimeOnAir}+\mathrm{RECEIVE\_DELAY}.
\end{equation}
From Eq.~\ref{eq:RTT} and Fig.~\ref{fig:timeonair} it can be seen that the
$\mathrm{RTT}$ delay of a message sent from the end-device to the controller,
and the response from the controller back to the end-device using LoRaWAN
ranges from $1.12$s to $6.59$s.
We do not include controller processing time in $\mathrm{RTT}$ because
$\mathrm{RECEIVE\_DELAY}$ is almost certainly larger than the processing time. 
An important observation to make is that $\mathrm{RTT}$ bounds the rate at
which events can be reported from the sensor to the controller.
An event that occurs less than $\mathrm{RTT}$ after a previous event will
be dropped due to the fact that the end-device will be awaiting a reply
from the controller.  

\subsection{Duty Cycle Limitation of LoRaWAN}

LoRaWAN operates in the unlicensed frequency band $863$-$870$MHz in Europe
(where this study was conducted).
European regulations impose duty cycles on users of this band to ensure fair
usage. 
When a LoRa end-device transmits a message, it can not use the same
channel for the length of its duty cycle.
The duty cycle ($\mathrm{DutyCycle}$) varies from $0.1$\% to $10$\% usage
time per node per channel.
For example, if an end-device spends $0.5$s transmitting a message on a
channel that specifies a $1$\% duty cycle, that channel will be unavailable
to the sending node for next $49.5$s.
Different channels can specify different duty cycles, as long as they remain
within the regulated duty cycle specifications.

We define the metric Blackout Period (BP) as
\begin{equation} \label{eq:toff}
\mathrm{BP} = \frac{\mathrm{TimeOnAir}}{\mathrm{DutyCycle}}-\mathrm{TimeOnAir}
\end{equation}
Blackout Period, $\mathrm{BP}$, is the time for which an end-device cannot
access a channel after it has sent a message due to duty cycle restriction.

We now relate $P_k^{i\rightarrow C}$, the event losses defined in Def.~2
of our ETC model, to the Blackout Period.
The cause of the BP, and therefore event losses, is the LoRaWAN duty cycle.  

An end-device might have $N$ channels available which decreases the blackout
time.
For example, if the end-device can transmit on $3$ channels instead of one,
each individual channel is still occupied for $1$\%.
However, the device is now transmitting for $1$\% of time units in each
channel, giving it a duty cycle of $3$\% which reduces $\mathrm{BP}$ in
Eq.~\ref{eq:toff}.
However, Eq.~\ref{eq:RTT} introduces the bound on minimum inter-arrival time of
two consecutive messages that has to be taken into account.
Therefore, the BP for an end-device that has the access to $N$ channels is given by
\begin{equation}
\mathrm{BP_N} = \mathrm{BP}-(N-1)\times\mathrm{RTT}.
\end{equation}

According to the LoRaWAN specification~\cite{LoRaWAN2015} a minimum of $3$
channels must be available to all end-devices.
An individual network may provide more channels.
For example, The Things Network \cite{TheThingsN} allocates $8$ channels with
$1$\% duty cycle to each end-device. 

\begin{figure}[!t]
\begin{center}
\includegraphics*[width=0.83\linewidth]{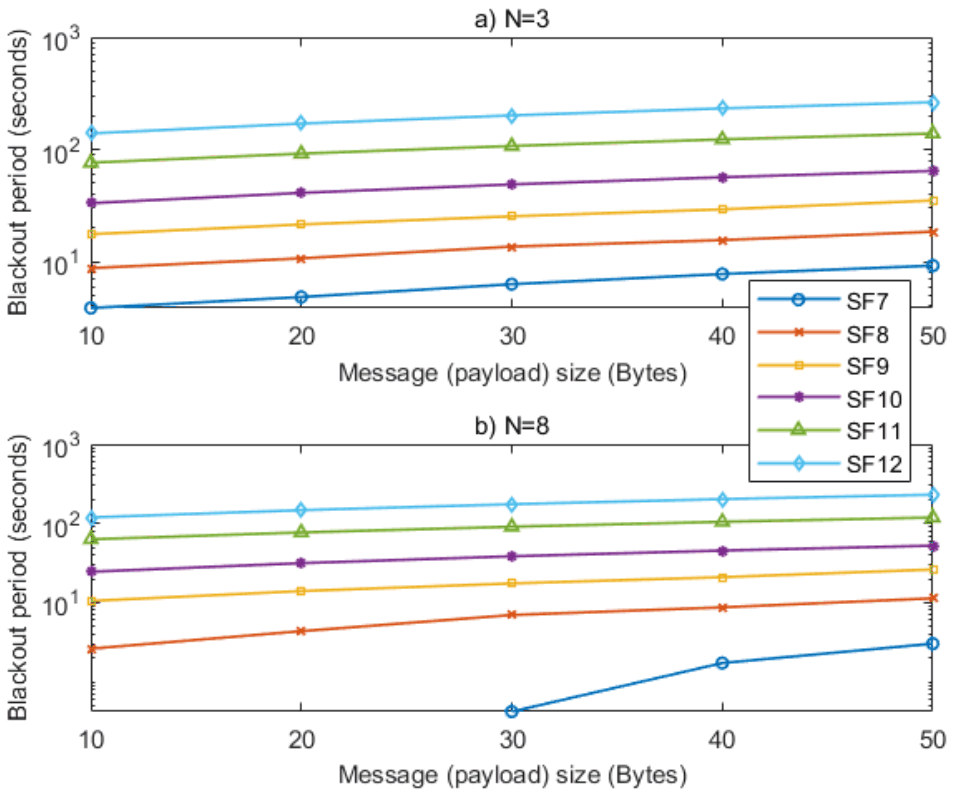}
\caption{\label{fig:blackouttime}Blackout Period of LoRaWAN for a) $N=3$ channels and b) $N=8$ channels, a code rate of $4/5$ and a bandwidth of $125$kHz (the y-axis uses logarithmic scale).}
\vspace{-5mm}
\end{center}
\end{figure}

We now examine the affect of multiple channels on the length of the BP.
Figure~\ref{fig:blackouttime}, part a) shows the BP of LoRaWAN end-device
when $3$ channels are available for various SFs and message sizes.
We can see that the BP varies from $3.8$s to $236.4$s when an end-device has
the access to only $3$ channels.
When the number of channels is increased to $8$, in Fig.~\ref{fig:blackouttime}, part b)
and a $1\%$ duty cycle is still maintained, the BP completely disappears for
SF$7$ payloads of $30$Bytes and less.
It is interesting to observe that the BP does not disappear for the other
SFs.
SF$10$, SF$11$, and SF$12$ only show a minimal improvement.
This shows that BPs can be reduced by adding more channels, but that it is
difficult to remove them completely, and that they get worse as the range
and robustness of the links improve. 

In the next section we will see the impact of the RTTs and the BPs on the
ability of ETC control schemes to maintain system stability and performance levels.
%
%
%
%
%

\section{Evaluation}
\label{sec:evaluation}

In this section, we present a simulation based experiment to illustrate the
feasibility of using LoRaWAN with an ETC scheme to maintain the stability
of a WNCS and guarantee a certain level of performance.
First, we present a dynamical model with an asynchronous ETC scheme.
The example is motivated by a real-life water network that can be characterised
as a slow-rate physical process.
Then, using projected Round Trip Time delays and Blackout Periods of data loss from Fig.~\ref{fig:timeonair} and Fig.~\ref{fig:blackouttime} we show that LoRaWAN
is capable of meeting the system's performance goal for certain scenarios.
Finally, we characterise the limitations of using LoRaWAN protocol if the 
application requirements increase or if the system is exposed to external
disturbances.

\subsection{Motivating Example}

Consider a hybrid linear model in the form of Eq.~\ref{eq:lin_sys} that
represents a water distribution network.
The state input vector $\xi(t) = [\xi_1(t) \enskip \xi_2(t) \enskip \xi_3(t)]^T$
represents the water levels in three different tanks.
Each tank is equipped with an end-device (sensor and actuator).
The model can work in two different modes for which the state matrix $A$ is
given as a zero matrix, i.e. $A_w=A_p=O_{3\times 3}$, and the input matrix
$B$ is given as
\small
\begin{align*}
B_w&=10^{-5} \times \left[\begin{array}{rrr}0.1436 & -0.0170 & -0.0164\\ -0.0098 & 0.1060 & -0.0100 \\ -0.0139 & -0.0139 & 0.1492\end{array}\right], \\
B_p&=10^{-5} \times \left[\begin{array}{rrr}0.7666 & -0.0493 & -0.0457\\ -0.0274 & 0.5848 & -0.0279 \\ -0.0393 & -0.0432 & 0.1492\end{array}\right].
\end{align*}
\normalsize
The control law in Eq.~\ref{eq:feedback_gain} works in a sample-and-hold fashion
with the feedback gain matrix $K$ given as
\small
\begin{align*}
K_w&=\left[\begin{array}{rrr}99950 & 3029 & 872\\ -3014 & 99940 & -1679 \\ -922 & 1652 & 99982\end{array}\right], \\
K_p&=\left[\begin{array}{rrr}9998.5 & 167.1 & 41.0\\ -166.6 & 9997.9 & -116.0 \\ -43.0 & 115.3 & 9999.2\end{array}\right].
\end{align*}
\normalsize
The \emph{'weak mode'} is represented by $(A_w, B_w, K_w)$ and it simulates
low water demand that a water network would experience during the night when
only the assistant pump is on.
The \emph{'powerful mode'} is represented by $(A_p, B_p, K_p)$ and it simulates
high water demand during the day when the powerful pump is enabled.

The model uses asynchronous ETC given in Sec.~\ref{sec:sys_model}.
The switching from powerful to weak mode is triggered by a function that maps
actuator saturation and quantization to represent the degree to which the
node's valve is open: $|S(-K_p\xi(t)+\alpha_p^{in})|_1<180^{\circ}$.
The valves themselves are discrete, and open and close in steps of $10^\circ$.
The term $|\cdot|_1$ is $\mathrm{L}^1$-norm, or sum of the entries, of the
resulting vector, and $\alpha_p^{in}$ denotes the degree that the in-valves
are open at equilibrium while in the powerful mode.
The switching from weak to powerful mode is triggered by $\xi_j(t)\leq h_{l_j}$
where $h_{l_j}$ is minimum water level for end-device $j$ and equals $0.03$m.
More details on the model can be found in \cite{Kartakis2017}.

\begin{figure}[!t]
\begin{center}
\includegraphics[width=0.85\linewidth]{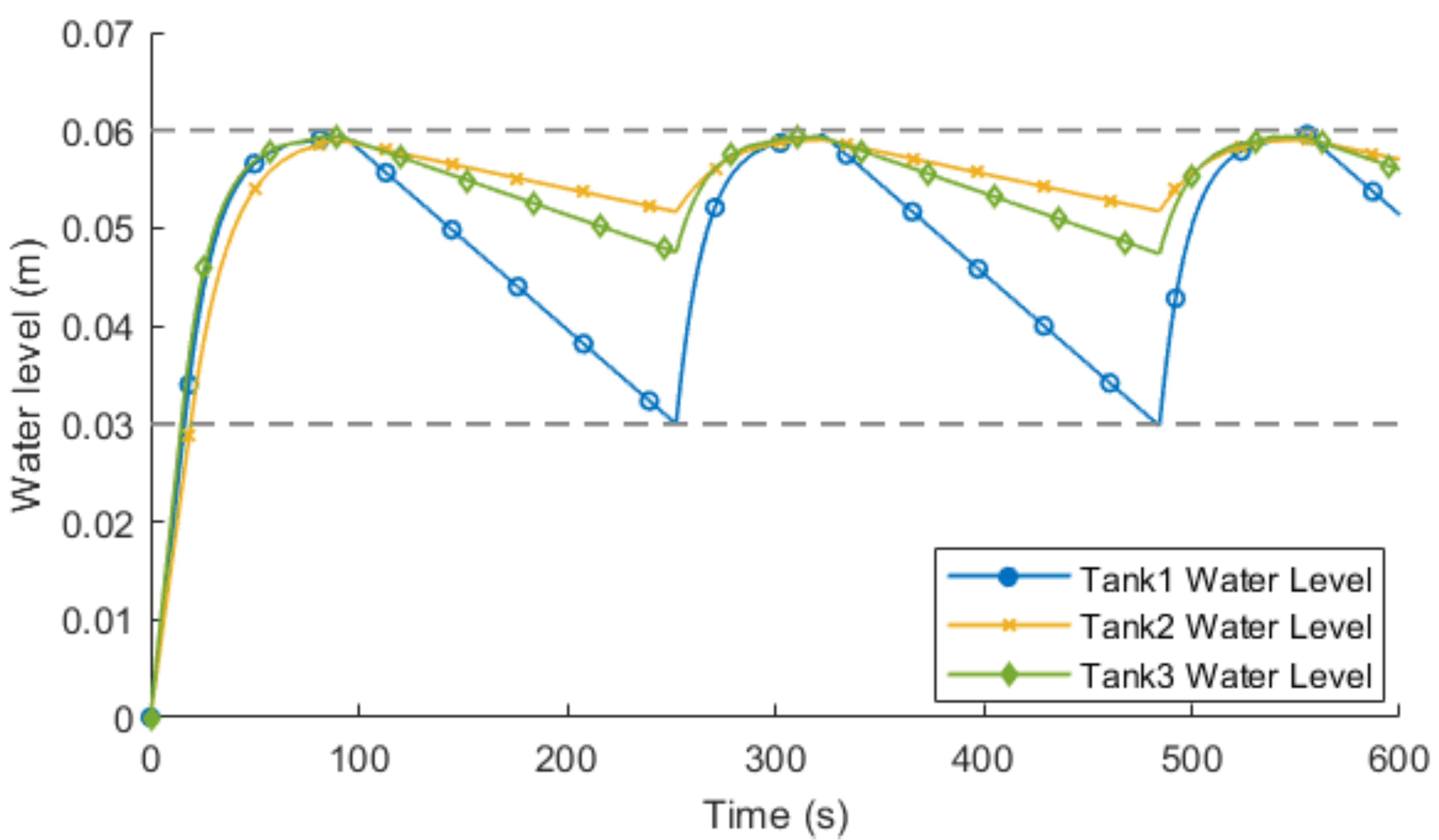}
\caption{\label{fig:normal_operation} The system response under normal operating conditions (no delays nor event losses).}
\vspace{-5mm}
\end{center}
\end{figure}

The system response under normal operating conditions is given in
Fig.~\ref{fig:normal_operation}.
It takes on average $100$ seconds for the system to reach the steady-state
value and continues operating within the safe bounds (in our case between $0.03$m
and $0.06$m).

\subsection{System Performance Analysis when using LoRaWAN}

We simulate the system for $600$s using a sampling rate of $1$ms.
Each end-device is allocated $N=3$ LoRaWAN channels with $1$\% duty
cycle restriction.
Our goal is to record overshoots and undershoots due to delays and
message losses caused by LoRaWAN limitations.
We measure this through the \emph{deviation metric} which indicates
the maximum system's deviation from the safe operating conditions which
is critical for water networks.
The behaviour of our system is understood and bounded.
Its acceptable deviation is up to $\pm15\%$.

\begin{figure}[!t]
\begin{center}
\includegraphics[width=0.85\linewidth]{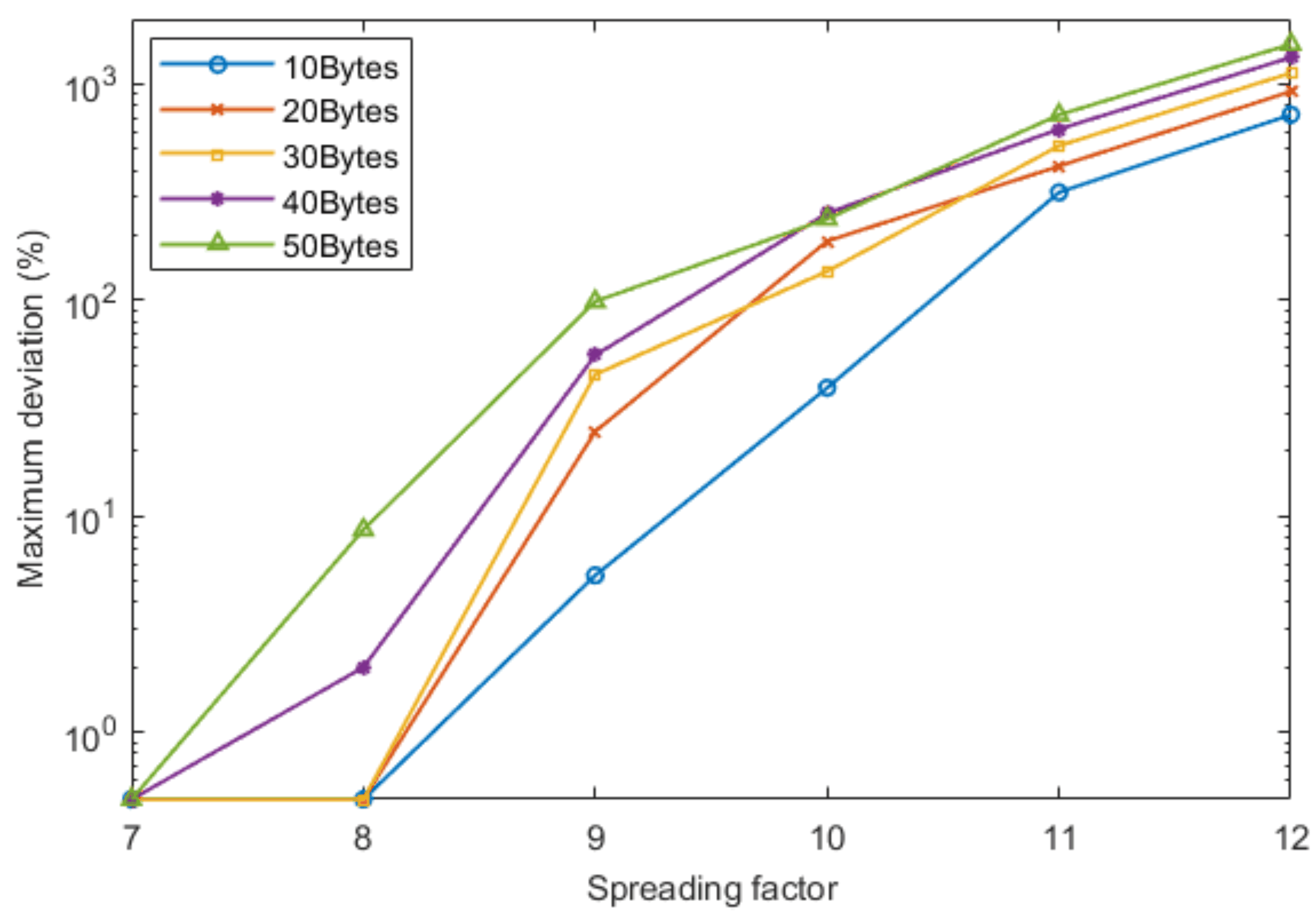}
\caption{\label{fig:max_dev} Maximum deviation of water level in the tank~$1$ when the communication is supported by LoRaWAN with $N=3$ channels (y-axis uses logarithmic scale).}
\vspace{-2mm}
\end{center}
\end{figure}

We consider two cases:

1) \emph{Disturbance-Free System} -
We assume an ideal case when there is no disturbance to the system.
We investigate the maximum system deviation for all spreading factors
(from $7$ to $12$) and various message sizes (from $10$Bytes to $50$Bytes
in increments of $10$).
The results are presented in Fig.~\ref{fig:max_dev}.
We present the results for the tank $1$ only as it is the smallest size tank and
therefore the most sensitive to state changes for identical control input.
As it can be observed from Fig.~\ref{fig:max_dev}, for low SFs and message sizes
the system is well within the bounds.
This is due to low RTTs and BPs as demonstrated in Sec.~\ref{sec:lora}.

2) \emph{System with Disturbance} - We model disturbance as a step impulse
in the water level state of the first tank, $\xi_1(t)$.
The magnitude of disturbance is equivalent to an increase in demand of $30\%$
at time $t$ or to a leak due to a pipe burst.
We vary the duration of the disturbance from $1$s to $10$s.
The results are presented in Table.~\ref{table:delays}.
These show that for severe disturbances (duration larger that $5$s), there
is a strict bound on SFs and message sizes that guarantee system performance
within desired levels.
To reject a disturbance, measurements need to be transmitted to the controller
in shorter time.
However, long BPs caused by LoRaWAN prevent reaction of the system at a rate that
can keep it stable.

\begin{table}[!t]
\caption{ Maximum deviation of water level (tank $1$) given in \% when there is an external step disturbance of $0.01$m ($N=3$ channels, SF8)}
\begin{center}
\begin{tabular}{l c c c}
\toprule
 \textbf{Disturbance Duration} & 1s & 5s & 10s \\ 
 \midrule
 \textbf{10Bytes} & 0.00 & 8.67 & 16.33 \\
 \textbf{20Bytes} & 1.67 & 9.23 & 18.00 \\
 \textbf{30Bytes} & 0.67 & 10.33 & 18.00 \\
 \textbf{40Bytes} & 0.67 & 10.33 & 22.67 \\
 \textbf{50Bytes} & 10.73 & 16.87 & 28.80 \\
 \bottomrule
\end{tabular} \vspace{-5mm}
\label{table:delays}
\end{center}
\end{table}

With this example we illustrated that LoRaWAN is able to support the classes
of applications where the underlying phenomenon changes at a slow rate.
The mode switching happens between every $50$s and $150$s.
LoRaWAN in such control system handles packet sizes of up to $50$Bytes for
SF7 and SF8 without affecting the system's performance and stability, as well
as the smallest packet size of up to $10$Bytes for SF7, SF8 and SF9.
LoRaWAN can be used for systems that are resilient to blackout period
s and delays of up to $10$s.
These limits become more conservative when the system is affected by significant
disturbances.
LoRaWAN protocol does not provide flexibility of sending time-critical data
if there is a duty cycle restriction in place at that moment.

\section{Conclusions} 
\label{sec:concl}

In this paper we created a model of a system using an ETC controller, and
a model of the LoRaWAN LPWA protocol.
We discussed the causes of delays and message losses in the LoRaWAN protocol,
and how these affect the stability and performance of the ETC controlled system.
Our results demonstrated the feasibility of the LoRaWAN protocol for certain
application scenarios and network parameter settings.
We also identified the limitations of LoRaWAN that prevent its wide adoption for use in certain classes of control-based application scenarios.

This paper represents a step forward in addressing the communication
and control systems co-design problem for long-range control over wireless
channel. 
It is an ongoing process and there is no single widely accepted methodology
to address the development of protocol stacks suitable for WNCSs. 
We showed that the current specification of LoRaWAN can be considered as
suitable for slow-changing physical processes where the data transmission
requirements are not high.
We will continue to address the coexistence of LPWA protocols and ETC control
by focusing on $n$ end-devices sharing the same set of channels.
We will also investigate the possibility of having priority slots, so even if
there is a duty cycle restriction in place there is a way to report a possible
failure due to the disturbance.
Additionally, we will exploit the possibility of using 'Listen before Talk'
feature of LoRa where by listening to the channel and sending if free the duty
cycle restriction can be bypassed.
\section*{Acknowledgment}

This work has been funded by the CISCO Research Centre/Silicon Valley Community Foundation, CG\# 955107, project 'Fog to FIELD: Securing Wide-area Monitoring and Control for Critical Infrastructure 4.0'.


%

\bibliographystyle{IEEEtran}
\bibliography{IEEEabrv,mybibfile}

\begin{thebibliography}{10}
\providecommand{\url}[1]{#1}
\csname url@samestyle\endcsname
\providecommand{\newblock}{\relax}
\providecommand{\bibinfo}[2]{#2}
\providecommand{\BIBentrySTDinterwordspacing}{\spaceskip=0pt\relax}
\providecommand{\BIBentryALTinterwordstretchfactor}{4}
\providecommand{\BIBentryALTinterwordspacing}{\spaceskip=\fontdimen2\font plus
\BIBentryALTinterwordstretchfactor\fontdimen3\font minus
  \fontdimen4\font\relax}
\providecommand{\BIBforeignlanguage}[2]{{%
\expandafter\ifx\csname l@#1\endcsname\relax
\typeout{** WARNING: IEEEtran.bst: No hyphenation pattern has been}%
\typeout{** loaded for the language `#1'. Using the pattern for}%
\typeout{** the default language instead.}%
\else
\language=\csname l@#1\endcsname
\fi
#2}}
\providecommand{\BIBdecl}{\relax}
\BIBdecl

\bibitem{Astrom1997}
K.~J. {{\AA}}str\"om and B.~Wittenmark, \emph{Computer-controlled Systems (3rd
  Ed.)}.\hskip 1em plus 0.5em minus 0.4em\relax Upper Saddle River, NJ, USA:
  Prentice-Hall, Inc., 1997.

\bibitem{Rajkumar2010}
R.~Rajkumar, I.~Lee, L.~Sha, and J.~Stankovic, ``Cyber-physical systems: The
  next computing revolution,'' in \emph{Proc. of the 47th Design Autom. Conf.},
  2010, pp. 731--736.

\bibitem{Ali2015}
S.~Ali, S.~B. Qaisar, H.~Saeed, M.~F. Khan, M.~Naeem, and A.~Anpalagan,
  ``Network challenges for cyber physical systems with tiny wireless devices: A
  case study on reliable pipeline condition monitoring,'' \emph{Sensors},
  vol.~15, no.~4, pp. 7172--7205, 2015.

\bibitem{Park2017}
P.~Park, S.~C. Ergen, C.~Fischione, C.~Lu, and K.~H. Johansson, ``Wireless
  network design for control systems: A survey,'' \emph{IEEE Commun. Surveys
  Tuts.}, vol.~PP, no.~99, pp. 1--1, 2017.

\bibitem{Wu2017}
B.~Wu, M.~D. Lemmon, and H.~Lin, ``Formal methods for stability analysis of
  networked control systems with ieee 802.15.4 protocol,'' \emph{IEEE Trans. on
  Control Sys. Tech.}, vol.~PP, no.~99, pp. 1--11, 2017.

\bibitem{Tabuada2007}
P.~Tabuada, ``Event-triggered real-time scheduling of stabilizing control
  tasks,'' \emph{IEEE Trans. on Autom. Control}, vol.~52, no.~9, pp.
  1680--1685, 2007.

\bibitem{Raza2017}
U.~Raza, P.~Kulkarni, and M.~Sooriyabandara, ``Low power wide area networks: An
  overview,'' \emph{IEEE Commun. Surveys Tuts.}, vol.~19, no.~2, pp. 855--873,
  2017.

\bibitem{LoRaWAN2015}
N.~Sornin, M.~Luis, T.~Eirich, T.~Kramp, and O.~Hersent, ``{LoRaWAN
  Specification, January 2015}.''

\bibitem{Wang2011}
X.~Wang and M.~D. Lemmon, ``Event-triggering in distributed networked control
  systems,'' \emph{IEEE Trans. on Autom. Control}, vol.~56, no.~3, pp.
  586--601, 2011.

\bibitem{Lehmann2012}
D.~Lehmann and J.~Lunze, ``Event-based control with communication delays and
  packet losses,'' \emph{Int. Journal of Control}, vol.~85, no.~5, pp.
  563--577, 2012.

\bibitem{Mazo2014}
M.~Mazo and M.~Cao, ``Asynchronous decentralized event-triggered control,''
  \emph{Automatica}, vol.~50, no.~12, pp. 3197 -- 3203, 2014.

\bibitem{Guinaldo2012}
M.~Guinaldo, D.~Lehmann, J.~Sánchez, S.~Dormido, and K.~H. Johansson,
  ``Distributed event-triggered control with network delays and packet
  losses,'' in \emph{51st IEEE Conf. on Decision and Control}, 2012, pp. 1--6.

\bibitem{cattani2017experimental}
M.~Cattani, C.~A. Boano, and K.~R{\"o}mer, ``An experimental evaluation of the
  reliability of lora long-range low-power wireless communication,''
  \emph{Journal of Sensor and Actuator Networks}, vol.~6, no.~2, p.~7, 2017.

\bibitem{Gutierrez2001}
J.~A. Gutierrez, M.~Naeve, E.~Callaway, M.~Bourgeois, V.~Mitter, and B.~Heile,
  ``Ieee 802.15.4: a developing standard for low-power low-cost wireless
  personal area networks,'' \emph{IEEE Network}, vol.~15, no.~5, pp. 12--19,
  2001.

\bibitem{TheThingsN}
\BIBentryALTinterwordspacing
(2018) {The Things Network}. [Online]. Available:
  \url{https://www.thethingsnetwork.org/}
\BIBentrySTDinterwordspacing

\bibitem{Kartakis2017}
S.~Kartakis, A.~Fu, M.~Mazo, and J.~A. McCann, ``Communication schemes for
  centralized and decentralized event-triggered control systems,'' \emph{IEEE
  Trans. on Control Syst. Technol.}, pp. 1--14, 2017.

\end{thebibliography}




\begin{IEEEbiography}{Michael Shell}
Biography text here.
\end{IEEEbiography}

\begin{IEEEbiographynophoto}{John Doe}
Biography text here.
\end{IEEEbiographynophoto}

\begin{IEEEbiographynophoto}{Jane Doe}
Biography text here.
\end{IEEEbiographynophoto}

\end{document}